\documentclass[useAMS,usenatbib]{mn2e}
\usepackage{aas_macros,graphicx,times,multirow,amsmath,amssymb,color,longtable}
\title[Discovery of the eclipsing binary \src]{Swift\,J201424.9+152930: discovery of a new deeply eclipsing binary with 491-s and 3.4-h modulations}
\author[P.~Esposito et al.] {P.~Esposito,$^{1,2}$\thanks{E-mail: paoloesp@iasf-milano.inaf.it} G.~L.~Israel,$^{3}$  D.~de~Martino,$^4$ P.~D'Avanzo,$^{5}$ V.~Testa,$^{3}$  L.~Sidoli,$^{1}$ \newauthor  R.~Di~Stefano,$^2$ A.~Belfiore,$^{1}$ M. Mapelli,$^6$ S.~Piranomonte,$^{3}$
 G.~A.~Rodr\'iguez~Castillo,$^{3}$ \newauthor  A.~Moretti,$^{7}$ V.~D'Elia,$^{3,8}$ F.~Verrecchia,$^{3,8}$ S.~Campana$^{5}$ and N.~Rea$^{9,10}$
\smallskip\\
$^1$Istituto di Astrofisica Spaziale e Fisica Cosmica - Milano, INAF, via E. Bassini 15, I-20133 Milano, Italy\\
$^2$Harvard--Smithsonian Center for Astrophysics, 60 Garden Street, Cambridge, MA 02138, USA\\
$^3$Osservatorio Astronomico di Roma, INAF, via Frascati 33, I-00040 Monteporzio Catone, Italy\\
$^4$Osservatorio Astronomico di Capodimonte, INAF, salita Moiariello 16, I-80131 Napoli, Italy\\
$^5$Osservatorio Astronomico di Brera, INAF, via E. Bianchi 46, I-23807, Merate (LC), Italy\\
$^6$Osservatorio Astronomico di Padova, INAF, vicolo dell'Osservatorio 5, I-35122, Padova, Italy\\
$^7$Osservatorio Astronomico di Brera, INAF, via Brera 28, I-20121 Milano, Italy\\
$^8$ASI Science Data Center (ASDC), via del Politecnico snc, I-00133 Roma, Italy\\
$^9$Anton Pannekoek Institute, University of Amsterdam, Postbus 94249, 1090GE Amsterdam, The Netherlands\\
$^{10}$Institut de Ci\`encies de l'Espai (CSIC-IEEC), Campus UAB, Facultat de Ci\`encies, Torre C5-parell, E-08193 Barcelona, Spain}
\date{Accepted 2015 March 27.  Received 2015 March 25; in original form 2014 December 9} \pagerange{\pageref{firstpage}--\pageref{lastpage}} \pubyear{2015}

\def\LaTeX{L\kern-.36em\raise.3ex\hbox{a}\kern-.15em
    T\kern-.1667em\lower.7ex\hbox{E}\kern-.125emX}
\def\xmm {\emph{XMM--Newton}}
\def\cxo {\emph{Chandra}}
\def\swift {\emph{Swift}}

\def\src {Sw\,J2014}
\def\flux {\mbox{erg cm$^{-2}$ s$^{-1}$}}
\def\lum {\mbox{erg s$^{-1}$}}
\def\nh {$N_{\rm H}$}

\begin{document}

\label{firstpage}
\maketitle
\begin{abstract}
We report on the discovery of a new X-ray pulsator, Swift\,J201424.9+152930 (\src). Owing to its X-ray modulation at 491~s, it was discovered in a systematic search for coherent signals in the archival data of the \swift\ X-ray Telescope. To investigate the nature of \src, we performed multi-wavelength follow-up observations with space-borne (\swift\ and \xmm) and ground-based (the 1.5-m Loiano Telescope and the 3.6-m Telescopio Nazionale Galileo) instruments. The X-ray spectrum of \src\ can be described by a hard and highly absorbed ($N_{\rm H}\sim5\times10^{22}$~cm$^{-2}$) power law ($\Gamma \sim 1$). The optical observations made it possible to single out the optical counterpart to this source, which displays several variable emission lines and total eclipses lasting $\approx$20 min. Total eclipses of similar length were observed also in X-rays. The study of the eclipses, allowed us to infer a second periodicity of 3.44~h, which we interpret as the orbital period of a close binary system. We also found that the period has not significantly changed over a $\sim$7~yr timespan. 
Based on the timing signatures of \src, and its optical and X-ray spectral properties, we suggest that it is a close binary hosting an accreting magnetic white dwarf. The system is therefore a cataclysmic variable of the intermediate polar type and one of the very few showing deep eclipses.
\end{abstract}

\section{Introduction}

Temporal investigation is crucial to properly identify the nature of an X-ray source or, at least, to significantly constrain it. For example, the discovery of type I (thermonuclear) X-ray bursts in a binary system securely identifies the compact object as a neutron star (NS), while the detection of fast X-ray pulsations rules out a black hole. In particular, timing analysis aimed at detecting periodic signals is one of the most powerful tools to pin down the nature of a source. For this reason, we have been working for years at systematic searches for coherent or quasi-coherent signals in the archival data of the \cxo\ \citep{weisskopf02} and \swift\ \citep{gehrels04} missions.\footnote{We are about to start a similar search in the X-ray archival data of \xmm\ \citep{jansen01}. This effort will be undertaken in the framework of the EXTraS project (Exploring the X-ray Transient and variable Sky, see http://www.extras-fp7.eu/ for details). } These projects, the \cxo\ ACIS Timing Survey at Brera And Rome astronomical observatories (CATS\,@\,BAR) and the \swift\ Automatic Timing Survey at Brera And Roma astronomical observatories (SATS\,@\,BAR, formerly known as SATANASS\,@\,BAR)  have led so far to the discovery of about 40 new X-ray periodicities. Also taking advantage of optical observations, we have been able to securely classify the nature of several of these pulsators (see \citealt{eis13,eisrc13,eism13,eis14}).

In this paper we present a detailed study of the data of the X-ray point source Swift\,J201424.9+152930 (\src), which was serendipitously imaged during \swift\ followup observations of the gamma-ray burst GRB\,061122 \citep{gotz13}. During a run of the SATS\,@\,BAR pipeline (see Section\,\ref{timing} for more details), we found a statistically significant periodicity in its X-ray emission, at $\sim$491~s. To investigate the nature of \src, we obtained further \swift\ and new \xmm\ data, as well as optical (photometric and spectroscopic) observations with the Cassini telescope in Loiano and the Telescopio Nazionale Galileo (TNG) telescope in La Palma. In Sections \ref{xobs}, \ref{tngsect}, and \ref{loiano} we present the data and the results from their analysis, in Section\,\ref{discussion} we discuss the nature of the source. A summary follows in Section\,\ref{summary}.

\section{\emph{Swift} and \emph{XMM--Newton}}\label{xobs}

\subsection{\swift\ observations and data reduction}

\src\ was discovered in a series of 14 \swift\ observations devoted to GRB\,061122 (target ID 20044), which were carried out in 2006 November through December. In these pointings, the source was generally $\sim$11$'$ from the  aim point of the X-Ray Telescope (XRT; \citealt{burrows05}) and the average exposure was $\sim$12.8~ks. In Table\,\ref{xrt}, we give a summary of these \swift\ observations, except for observations 4001 (where \src\ was outside the field of view) and 4009 (where it was at the very edge of the detector), which were not used for this work.
Additional \swift\ observations pointed at \src\ (target ID 46265) were performed between 2011 December and 2013 December (see Table\,\ref{xrt}). Most of these observations were taken at the end of 2013, and have typical exposure of 9--10~ks each.

Each XRT data set was calibrated, filtered and screened according to standard criteria using \textsc{xrtpipeline} within the \textsc{xrtdas} package, included in the \textsc{HEAsoft} distribution v6.15. At the same time, exposure maps were produced for all observations.
The X-ray source counts for the spectral and timing analysis were accumulated in circles centred on \src\ and with radii from $\sim$30$''$ to 47$''$, depending on the off-axis distance, while the background was estimated from neighbouring, source-free regions. Since in the individual observations the count statistics were not enough to allow us a reasonable number of spectral bins, we created two combined spectra: one by merging the 2006 observations (total exposure 153.0~ks, about 1000 net source counts) and one from the 2013 November--December observations (5011--5037, total exposure 169.1~ks, about 1100 net source counts).
\begin{table*}
\begin{minipage}{16cm}
\centering \caption{Summary of the \swift\ and \xmm\ X-ray observations of \src.}\label{xrt}
\begin{tabular}{@{}lcccc}
\hline
Obs.\,ID  & Start & Stop & Net exposure & Count rate$^{a}$ \\
 & \multicolumn{2}{c}{(YYYY-MM-DD hh-mm-ss)} & (ks) & (counts~s$^{-1}$) \\
\hline
\swift/XRT~00020044002 & 2006-11-23 01:37:55 & 2006-11-24 22:47:56 & 16.0 & $(9.5\pm1.4)\times10^{-3}$ \\
\swift/XRT~00020044003 & 2006-11-25 00:14:53 & 2006-11-25 22:58:58 & 14.2 & $(6.9\pm1.2)\times10^{-3}$ \\
\swift/XRT~00020044004 & 2006-11-26 00:22:02 & 2006-11-26 22:57:56 & 10.6 & $(10.7\pm1.9)\times10^{-3}$ \\
\swift/XRT~00020044005 & 2006-11-27 00:27:53 & 2006-11-27 23:15:57 & 12.9 & $(9.7\pm1.5)\times10^{-3}$ \\
\swift/XRT~00020044006 & 2006-11-28 00:32:54 & 2006-11-28 23:16:57 & 15.5 & $(4.5\pm0.9)\times10^{-3}$ \\
\swift/XRT~00020044007 & 2006-11-29 00:37:57 & 2006-11-29 23:25:58 & 16.0 & $(11.9\pm1.3)\times10^{-3}$ \\
\swift/XRT~00020044008 & 2006-11-30 00:45:18 & 2006-11-30 23:37:56 & 17.0 & $(11.8\pm1.3)\times10^{-3}$ \\
\swift/XRT~00020044010 & 2006-12-02 10:34:55 & 2006-12-03 02:46:38 & 7.8 & $(10.1\pm1.8)\times10^{-3}$ \\
\swift/XRT~00020044011 & 2006-12-03 04:14:54 & 2006-12-04 01:31:25 & 8.7 & $(1.3\pm0.2)\times10^{-2}$ \\
\swift/XRT~00020044012 & 2006-12-04 05:57:54 & 2006-12-04 23:59:56 & 10.7 & $(11.2\pm1.6)\times10^{-3}$ \\
\swift/XRT~00020044013 & 2006-12-05 01:18:07 & 2006-12-05 23:59:56 & 13.2 & $(1.6\pm0.2)\times10^{-2}$ \\
\swift/XRT~00020044014 & 2006-12-06 01:20:57 & 2006-12-06 23:59:56 & 11.1 & $(11.1\pm1.6)\times10^{-3}$ \\
\swift/XRT~00046265001 & 2011-12-26 21:01:46 & 2011-12-26 21:06:59 & 0.3 & $<$$4.8\times10^{-2}$ \\
\swift/XRT~00046265002 & 2012-08-27 09:04:28 & 2012-08-27 09:08:53 & 0.3 & $<$$6.2\times10^{-2}$ \\
\swift/XRT~00046265003 & 2012-09-24 21:42:36 & 2012-09-24 23:28:54 & 1.0 & $(1.1\pm0.4)\times10^{-2}$ \\
\swift/XRT~00046265004 & 2012-12-26 01:51:34 & 2012-12-26 23:10:55 & 3.9 & $(4.7\pm1.5)\times10^{-3}$ \\
\swift/XRT~00046265005 & 2013-03-08 04:52:49 & 2013-03-08 11:16:54 & 0.2 & $<$$6.3\times10^{-2}$ \\ 
\swift/XRT~00046265007 & 2013-03-15 00:24:59 & 2013-03-16 13:24:55 & 8.7 & $(2.0\pm0.2)\times10^{-2}$ \\
\swift/XRT~00046265008 & 2013-03-18 09:55:58 & 2013-03-18 15:07:56 & 1.8 & $(4.9\pm2.3)\times10^{-3}$ \\
\swift/XRT~00046265009 & 2013-07-02 12:41:02 & 2013-07-02 12:45:55 & 0.3 & $<$$5.2\times10^{-2}$ \\
\swift/XRT~00046265010 & 2013-09-16 02:11:05 & 2013-09-16 02:17:53 & 0.4 & $<$$2.8\times10^{-2}$ \\ 
\swift/XRT~00046265011 & 2013-11-27 01:51:25 & 2013-11-27 22:51:55 & 8.9 & $(14.0\pm1.7)\times10^{-3}$ \\
\swift/XRT~00046265012 & 2013-11-28 00:12:26 & 2013-11-28 10:19:55 & 9.7 & $(15.8\pm1.7)\times10^{-3}$ \\
\swift/XRT~00046265013 & 2013-11-29 00:14:40 & 2013-11-29 14:46:56 & 9.5 & $(9.6\pm1.5)\times10^{-3}$ \\
\swift/XRT~00046265014 & 2013-11-30 00:15:40 & 2013-11-30 14:46:56 & 8.6 & $(3.8\pm0.9)\times10^{-3}$ \\
\swift/XRT~00046265015 & 2013-12-01 00:14:30 & 2013-12-01 14:48:56 & 9.6 & $(5.3\pm1.0)\times10^{-3}$ \\
\swift/XRT~00046265016 & 2013-12-02 00:15:02 & 2013-12-02 14:47:54 & 9.7 & $(6.1\pm1.1)\times10^{-3}$ \\
\swift/XRT~00046265018 & 2013-12-03 01:53:39 & 2013-12-03 16:26:57 & 9.4 & $(6.4\pm1.1)\times10^{-3}$ \\
\swift/XRT~00046265020 & 2013-12-04 00:17:55 & 2013-12-04 16:34:54 & 9.8 & $(9.6\pm1.4)\times10^{-3}$ \\
\swift/XRT~00046265024 & 2013-12-06 00:20:39 & 2013-12-06 12:05:54 & 9.6 & $(15.5\pm1.7)\times10^{-3}$ \\
\swift/XRT~00046265025 & 2013-12-07 00:23:36 & 2013-12-07 11:49:56 & 9.7 & $(5.4\pm1.0)\times10^{-3}$ \\
\swift/XRT~00046265027 & 2013-12-08 00:21:12 & 2013-12-08 16:25:55 & 9.6 & $(3.3\pm0.8)\times10^{-3}$ \\
\swift/XRT~00046265028 & 2013-12-09 00:26:36 & 2013-12-09 13:29:54 & 9.6 & $(5.9\pm1.1)\times10^{-3}$ \\
\swift/XRT~00046265030 & 2013-12-10 08:25:39 & 2013-12-10 22:57:54 & 9.9 & $(4.0\pm0.8)\times10^{-3}$ \\
\swift/XRT~00046265032 & 2013-12-11 03:38:36 & 2013-12-11 19:49:55 & 9.5 & $(5.6\pm1.1)\times10^{-3}$ \\
\swift/XRT~00046265034 & 2013-12-12 00:32:54 & 2013-12-12 18:14:56 & 9.4 & $(6.1\pm1.1)\times10^{-3}$ \\
\swift/XRT~00046265036 & 2013-12-13 00:30:44 & 2013-12-13 19:57:54 & 9.3 & $(9.2\pm1.4)\times10^{-3}$ \\
\swift/XRT~00046265037 & 2013-12-14 00:32:04 & 2013-12-14 23:09:55 & 9.3 & $(16.6\pm1.9)\times10^{-3}$ \\
\hline
\xmm/EPIC~0743980501 (pn) & 2014-05-02 11:58:12 & 2014-05-02 17:25:23 & 16.7 & $0.105\pm0.005$\\
\xmm/EPIC~0743980501 (MOS\,1) & 2014-05-02 11:35:27 & 2014-05-02 17:20:43 & 20.5 & $0.022\pm0.002$ \\
\xmm/EPIC~0743980501 (MOS\,2) & 2014-05-02 11:36:00 & 2014-05-02 17:20:55 & 20.5 & $0.032\pm0.002$ \\
\hline
\end{tabular}
\begin{list}{}{}
\item[$^{a}$] Net source count rate in the 0.3--10~keV band; when the source is not detected, a 3$\sigma$ upper limit is given.
\end{list}
\end{minipage}
\end{table*}

\src\ was in the Ultra-Violet/Optical Telescope (UVOT; \citealt{roming05}) field of view only in the observations pointed to the source (target ID 46265). The data were collected in imaging mode in near ultraviolet (UV), with the filters $uw1$ (central wavelength 2600~\AA, FWHM 693~\AA),  $um2$ (2246~\AA, FWHM 498~\AA), and $uw2$ (1928~\AA, FWHM 657~\AA).
The analysis was performed for each filter on the stacked images of the individual observations with the \textsc{uvotsource} task, which calculates count rates, flux densities and magnitudes through aperture photometry within a circular region and applies specific corrections due to the detector characteristics. Since \src\ is in a crowded field (see Section\,\ref{loiano}), we adopted an extraction radius of 3~arcsec and applied the corresponding aperture corrections.

\subsection{\xmm\ observations and data reduction}

\xmm\ observed \src\ on 2014 May 02 for approximately 21~ks (Obs. ID: 0743980501). The EPIC pn \citep{struder01} and MOS \citep{turner01} detectors were set in Full Frame mode and used the Medium optical blocking filter. The OM telescope \citep{mason01} was operated in Fast mode and the $B$ filter (390--490~nm) was used during the observation for all exposures. The exposure time was 19.6~ks for the pn (16.7~ks net exposure) and 20.7~ks for the MOS\,1 and MOS\,2 (20.5~ks net exposure).  Seven OM exposures were scheduled, of duration 1.0, 4.4, 4.4, 4.0, 1.2, 4.0, and 1.2~ks; however, due to a problem with the data processing unit (DPU), the OM data of the first three exposures were lost and the OM data started $\sim$10~ks after the EPIC ones, for a total exposure time of $\sim$10.4~ks. 

The raw observation EPIC data files were processed with the \xmm\ Science Analysis Software (\textsc{sas}, v.13.5) using standard pipeline tasks, \textsc{epproc} and \textsc{emproc} for the pn and MOSs, respectively. Likewise, the OM data were processed with the \textsc{sas} \textsc{omfchain} pipeline and the \src's  light curves were binned in 20-s intervals. Standard screening criteria were applied. 

The X-ray data were affected by very strong soft-proton flares, in particular during the second half of the exposure. For the spectral analysis, we removed the intervals of flaring background with a 3$\sigma$ clipping filter. This resulted in a net exposure time of 7.8~ks in the pn, 11.3~ks in the MOS\,1, and 11.5~ks in the MOS\,2.  For the timing analysis, since the same procedure would significantly alter the time series, we choose instead to dynamically subtract the scaled background in each bin of the source light curves binned at 10~s.

\subsection{X-ray position}
\begin{figure}
\centering
\resizebox{\hsize}{!}{\includegraphics[angle=-90]{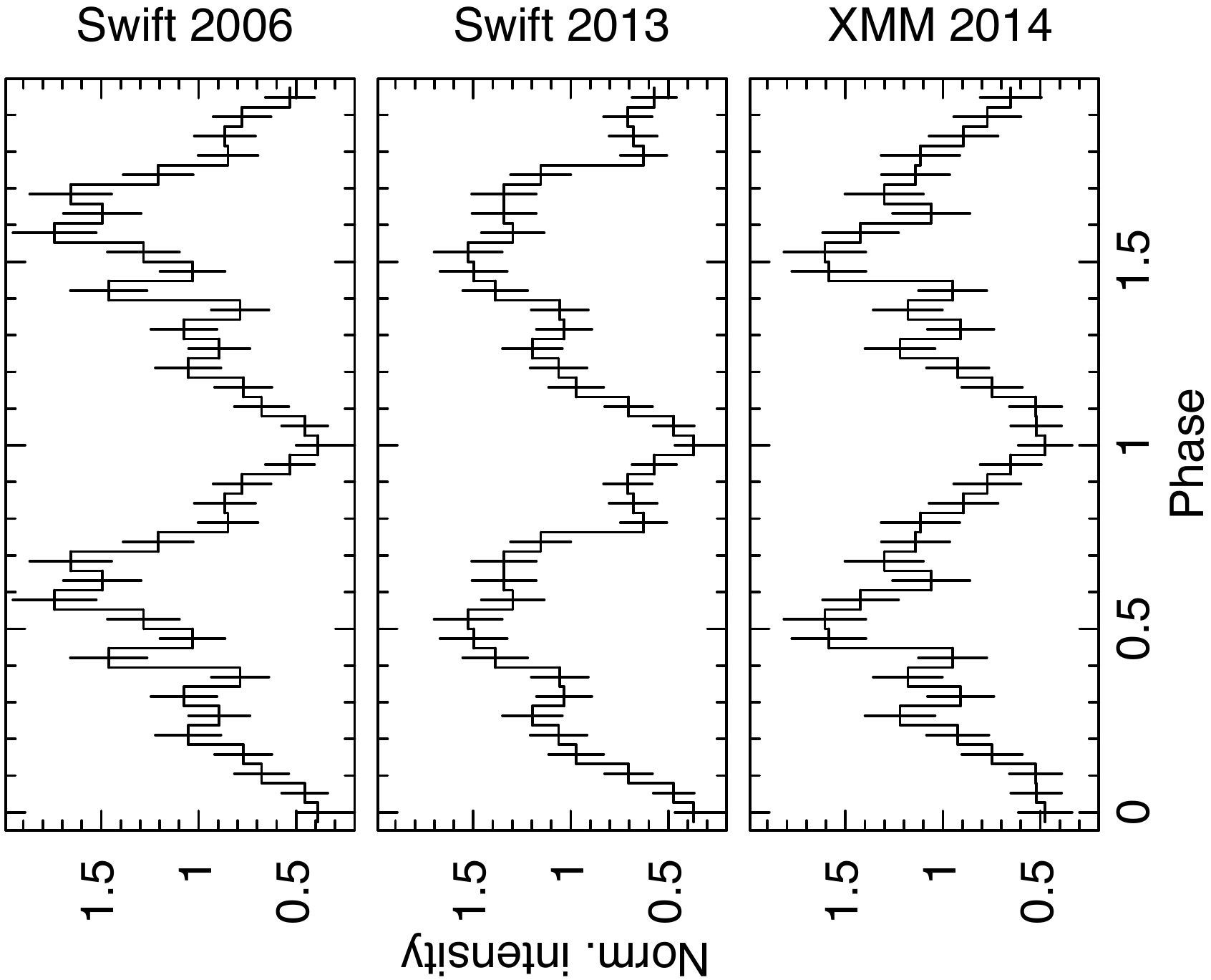}}
\caption{\label{efold_allX} \swift\ and \xmm\ folded light curves (491-s modulation, background-subtracted). Each data set was folded at its best period and the profiles have been aligned so that the minimum bin coincides with phase 0.}
\end{figure}
In X-rays, due to the larger number of collected photons, \xmm\ offers much better positional accuracy with respect to \swift/XRT.
Based on the combined pn and MOS images, the \textsc{emldetect} routine reports for \src\ the best-fitting position (J2000) $\rm RA=20^h14^m24\fs96,~Decl.=+15\degr29'29\farcs04$, with a statistical uncertainty of $0\farcs2$ (radius; here and in the following all uncertainties are at 1$\sigma$ confidence level). This is much smaller than the absolute astrometric accuracy of \xmm, $1\farcs5$ root mean square \citep{kirsch04}, which is thus a more realistic indication of the actual uncertainty. About 20 other weak X-ray sources are present in the EPIC images (they are all fainter than \src\ and none is closer than 2.4 arcmin to it) but, since they have no obvious and unambiguous optical or infrared counterparts, it was not possible to use them to improve the \xmm\ absolute astrometry.

 In all the optical data that we will discuss in the following, there is always only a single source consistent with the X-ray position of \src. Thus, we identify that source (see Section\,\ref{loiano}) as the counterpart of \src. Finally, we note that this optical counterpart is consistent with the NOMAD\footnote{The Naval Observatory Merged Astrometric Dataset (NOMAD; \citealt{zacharias05}) of the United States Naval Observatory (USNO), see http://www.usno.navy.mil/USNO/astrometry/optical-IR-prod/nomad.} source 1054-0592809 ($B\sim19.3$~mag, $R\sim18.9$~mag, 1050--16948097 in the USNO-A2 catalog and 1054--0559015 in the USNO-B1 catalogue, see also Section\,\ref{loiano}).

\subsection{Timing analysis}\label{timing}

The $\sim$491-s signal from \src\ was first detected within the SATS\,@\,BAR project. SATS\,@\,BAR consists in a systematic Fourier-based search for new pulsators in the archival \swift\ X-ray data. So far, about 4000 XRT light curves of point sources with a sufficiently high number of photons ($\ga$150) were analysed and the effort yielded six previously unknown X-ray pulsators (including \src, while three further new X-ray pulsators were reported in \citealt{nichelli09} and \citealt{eis14}). 

The original detection of the period of \src\ occurred in the piled \swift\ 2006 observations (see Table\,\ref{xrt}). The significance of the signal (as inferred from the automatic analysis and normalized for the number of searched frequencies and light curves) was above the a priori threshold of about 4$\sigma$. By a standard phase-fitting analysis, the value of the period was refined to $P=491.26(1)$~s.

After that, more \swift\ observations were carried out, in particular in late 2013. The period measured during the latest \swift\ campaign, using the observation from 2013 November 27 to December 14, $P=491.27(2)$~s, is consistent with that inferred in the 2006 observations. In our recent \xmm\ observation, executed on 2014 May 02, the period was $P=491(1)$~s, which is again consistent, albeit within a comparatively large error, with the previous measures. The 3$\sigma$ limit on the period derivative drawn from these measures by a linear fit is $-2.6\times10^{-10}~\mathrm{s~s}^{-1}<\dot{P}< 3.5\times10^{-10}~\mathrm{s~s}^{-1}$. The long-term stability of the period suggests that the periodic modulation is originated by the spin of a compact object, likely a white dwarf (WD; due to its larger momentum of inertia, a WD is much harder than an NS to spin-up or spin-down, see e.g. \citealt{patterson94,bildsten97,frank02}), or by the revolution of two rather close compact objects.  

The pulse profiles are shown in Fig.\,\ref{efold_allX}. The pulsed fractions, measured with a sine-wave fit,\footnote{The profiles were fitted with a model consisting of a constant, $C$, and a sine term, $f(\phi)= C +A\sin (\phi+\phi_0)$. We then determined the pulsed fraction as the amplitude of the
sine term $A$ divided by $C$.} were $48\pm5$\% in the 2006 \swift\ data, $50\pm5$\% in the 2013 \swift\ data, and $44\pm5$\% in the \xmm\ data. 
The data suggest a trend of pulsed fraction decreasing with energy. However, it is not statistically significant, since the values are compatible within the uncertainties. For example, in the \xmm\ data, the pulsed fraction is $47\pm8$\% in the soft ($<$3~keV) band and $36\pm4$\% in the hard ($>$3~keV) band. We also computed hardness ratios between profiles in a number of different energy bands (including the soft and hard band mentioned above), but no statistically significant deviation from the average ratio were observed along the 491-s cycle.
We searched also the \xmm/OM optical data for the 491-s periodicity by a Fourier transform, as well as by a folding analysis around the X-ray period. No statistically significant signal could be found. The 3$\sigma$ upper limit of the pulsed fraction, computed following \citet{vaughan94}, is 10\% for a sinusoidal signal. Finally, we notice that, although the pulse profiles (and the hardness ratios) obtained by folding the \swift\ and \xmm\ data of \src\ at the double of the period are rather symmetrical, we cannot exclude that the true period of the source is $\sim$982~s. 

\subsubsection{Eclipses and second (orbital) period}
\begin{figure*}
\centering
\resizebox{\hsize}{!}{\includegraphics[angle=-90]{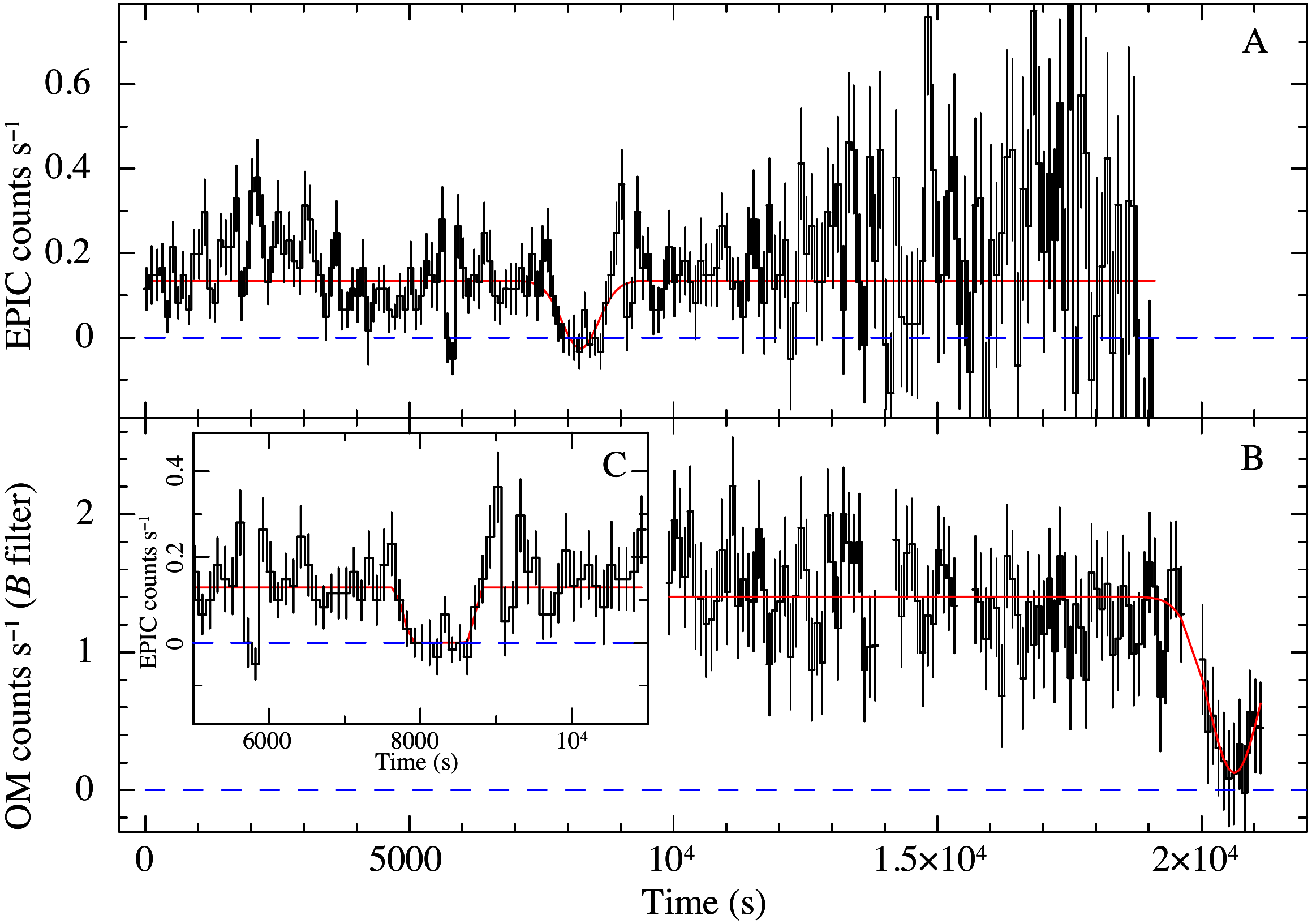}}
\caption{\label{xmmlcs} \xmm\ pn and OM light curves. In panels A and B, the red solid line indicates the constant-plus-Gaussian model used to infer the time and the duration of the eclipse. In inset C, we show a close-up of the X-ray eclipse, and the red line represents the piecewise linear model fit to the data.}
\end{figure*}

The inspection of the EPIC (pn + MOSs) and OM light curves clearly revealed two overall similar broad features, one in each instrument data set (Fig.\,\ref{xmmlcs}). In both instances, in a few hundreds of seconds the source flux suddenly decreased to a flux level consistent with zero and quickly recovered (to the pre-event flux level in the case of the EPIC data, while the last part of the optical eclipse was not observed by the OM). These properties are reminiscent of the total eclipses in low-mass X-ray binaries with an accreting NS or cataclysmic variables (CVs).

To determine the epoch and duration of the eclipses, we modelled the data with Gaussian functions. The EPIC eclipse occurred at MJD 56\,779.6265(6) (Barycentric Dynamical Time (TDB), centre of the event) and could be modelled by a Gaussian with $\sigma=344\pm52$~s (FWHM duration:  $809\pm122$~s). The OM eclipse took place at MJD 56\,779.771(1) (TDB) and could be modelled by a Gaussian with $\sigma=493\pm141$~s (FWHM duration:  $1160\pm332$~s; but again we note that the OM coverage of the event is incomplete). The two events are separated by $\Delta t_{\mathrm{eclipses}}=12.39\pm0.15$~ks ($3.44\pm0.04$~h).
The \xmm\ data are too scant and noisy to allow one to resolve precisely the ingresses and egresses of the eclipses, but for the X-ray data we derived rough estimates fitting to the light curve a piecewise linear model (Fig.\,\ref{xmmlcs}, panel C). The ingress/egress times obtained in this way are of $0.18\pm0.08$~ks and the duration of the phase completely inside ingress and egress (second to third contact, the `flat-bottom' part) is $0.59\pm0.06$~ks (the total eclipse
duration, from first to fourth contact, is $\approx$0.95~ks). 

Unfortunately, we do not have simultaneous EPIC and OM coverage for any of the two events. In fact, due to a DPU problem, the OM science data of the first $\sim$10~ks of the observation, when the EPIC eclipse occurred, were lost. On the other hand, the EPIC instruments were switched-off well before the OM eclipse. 
This makes it more delicate to draw information on the geometry of the system. None the less, under the hypothesis that the eclipse is periodically recurring (note that we detected a similar event in the 2011 optical observation at the Loiano observatory, Section\,\ref{loiano}) and can be regarded as the signature of an orbital motion, we can test the two following hypothesis: (i) the X-ray and optical eclipses are simultaneous, X-ray and optical photons are emitted by the same region (or close regions) and with the same drop mechanism; in this case the time between the EPIC and OM events, about 12.4~ks, reflects the orbital period of the system, (ii) the X-ray and optical eclipses are not simultaneous, X-ray and optical emission are originated by two different areas and/or mechanisms. If the X-rays originate from the compact object and the optical from the companion, thus the time between the EPIC and OM events marks half orbital period duration.

\begin{figure}
\centering
\resizebox{\hsize}{!}{\includegraphics[angle=-90]{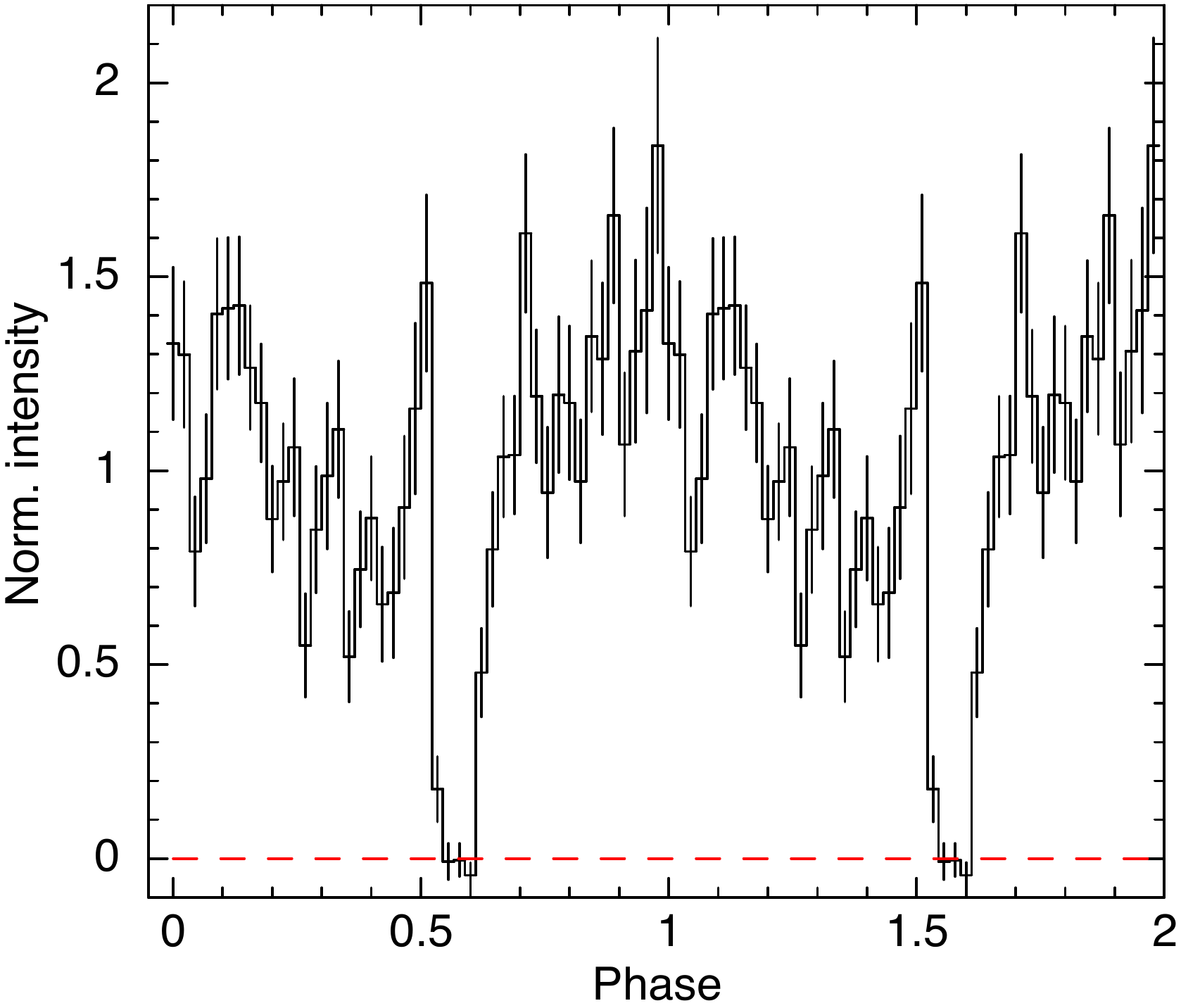}}
\caption{\label{swift_eclipse} Background-subtracted 0.3--10~keV  \swift/XRT light curve folded to the best inferred orbital period of 12\,379~s. The total eclipse, lasting about 1.2~ks, is clearly detected and in agreement with those displayed by the \xmm\ EPIC and OM instruments. }
\end{figure}

The \xmm\ data cannot disentangle the two scenarios. However, the \swift\ monitoring campaign carried out in 2006 and 2013 are better suited to this end. A period search with a Rayleigh periodogram around 12.4 and 24.8~ks (3 independent trials) shows two significant ($>$10$\sigma$) peaks at $12\,379\pm11$~s and $24\,758\pm20$~s (1$\sigma$ c.l. is used for the uncertainties).\footnote{The inspection of the power spectrum reveals the presence of another peak at a frequency of about $8\times10^{-5}$~Hz, which appears to be one of the sidelobes round the peak due to the \swift\ orbital revolution (at about 96~min, corresponding to $\sim$$1.7\times10^{-4}$~Hz). This suggests that the 12\,379-s signal was missed by the automatic SATS@BAR pipeline due to the presence of the nearby more intense peaks caused by the spacecraft's orbit.} 
The light curves folded at the two periods show a clear modulation with one eclipse in one cycle for the 12\,379~s period (Fig.\,\ref{swift_eclipse}), and two identical eclipses in one cycle for the 24\,758~s period. In both cases the eclipse duration is consistent with what observed with \xmm\ (see below). Correspondingly, we confidently believe that 12\,379~s is the correct periodicity and traces the orbital period of the system and thus, the periodicity at $\sim$491~s is the spin period of the compact object in \src. As a consequence, the X-ray and optical eclipses should occur simultaneously with photons emitted in the two bands originated by the same or very close regions. The epoch of the X-ray eclipses in the XRT data, with $P_{\mathrm{orb}}=12\,379$~s is MJD 56\,623.0287(18) TDB.

 The light curve folded at the orbital profile can be modelled with a sinusoidal modulation of amplitude (pulsed fraction) $26\pm3$\% with a superimposed eclipse of full width at half-maximum (FWHM) duration $1.2\pm0.1$~ks. The fit to eclipse of a piecewise linear model yields the following parameters: length of the flat-bottom part of $0.62\pm0.15$~ks and ingress/egress duration of $0.3\pm0.1$~ks.

\subsection{Spectral analysis}\label{spectra}

We tested a few simple models for the three \xmm/EPIC (pn + 2 MOSs, excluding the time of the eclipse) and \swift/XRT X-ray spectra of \src: power law, blackbody and thermal bremsstrahlung, all modified for the interstellar absorption. The spectra were grouped so as to have a minimum of 20 counts for energy channel, and the spectral fits were performed in the 1--10~keV energy band, owing to the very low signal-to-noise of the source data below 1~keV. All these models provide a satisfying fit to the data, but the temperature is totally unconstrained for the bremsstrahlung. For this reason, in Table\,\ref{spec} we summarize the results only for the power law and blackbody fits. 

\begin{figure}
\centering
\resizebox{\hsize}{!}{\includegraphics[angle=-90]{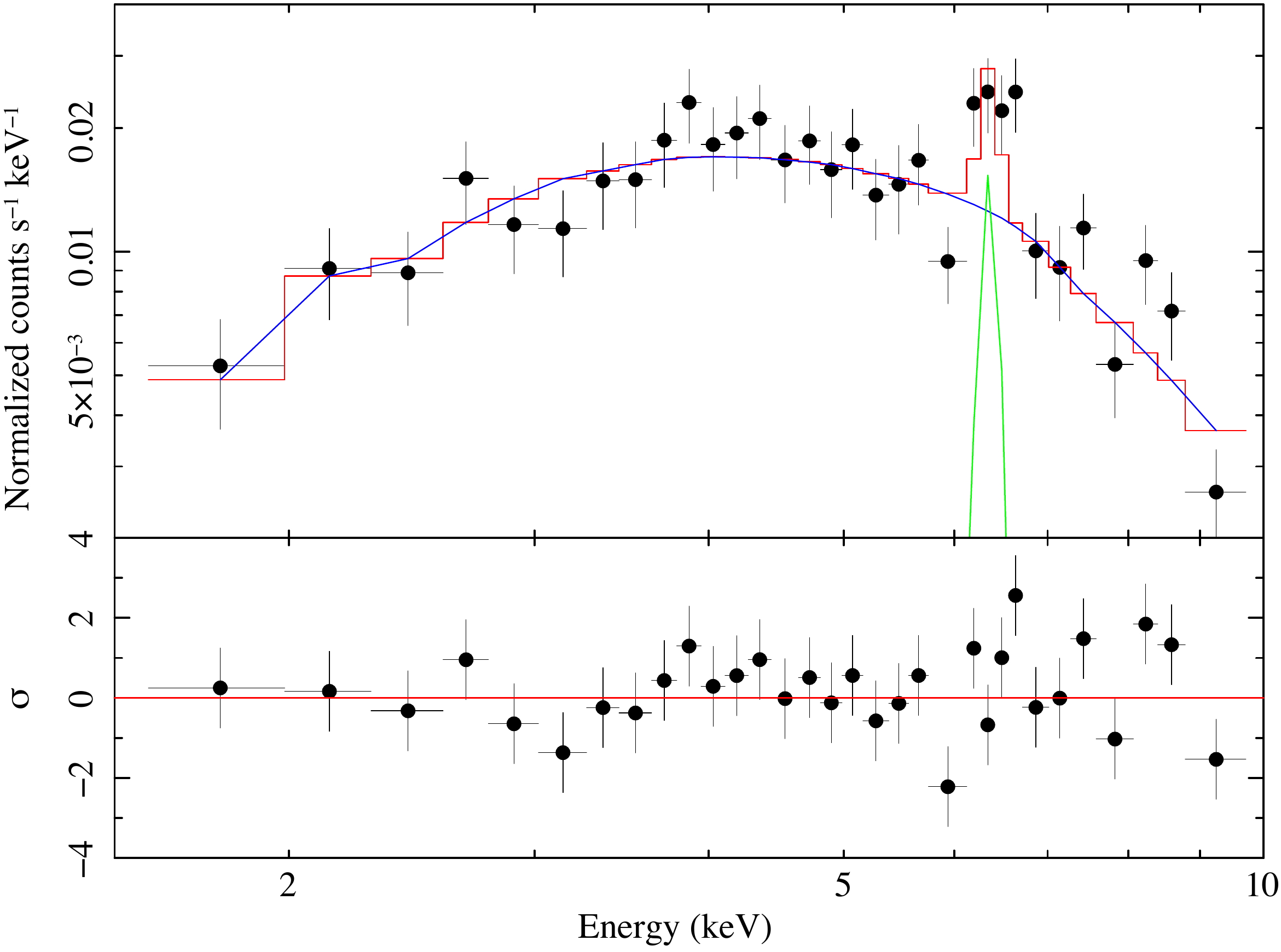}}
\caption{\label{xmmspec} \xmm\ pn spectrum and best-fitting model [\textsc{phabs(powerlaw+gaussian)}, in red]. The blue and green lines show the power-law and Gaussian component, respectively. Bottom panel: the residuals of the fit (in units of standard deviations).}
\end{figure}

The spectrum was rather hard in all three data sets, with a photon index $\Gamma\sim0.8$ (or temperature corresponding to $kT\sim2$~keV for the blackbody fits). The absorbing column inferred from the power-law fits is of $\sim$4--$5\times10^{22}$~cm$^{-2}$, which is much higher than the Galactic value in that direction ($\sim$$1.3\times10^{21}$~cm$^{-2}$).\footnote{From the \textsc{ftool nh} tool and based on the Galactic H\textsc{i} map by \citet{kalberla05}.} The measured fluxes show a moderate variability ($\sim$30\%) from the 2006 and 2013 \swift\ data to the 2004 \xmm\ observation. In the \xmm\ data, which are those with the best counting-statistics, the $\chi^2$ values improve with the addition of a Gaussian emission feature to model an excess around 6.5~keV which is probably due to Fe lines (Fig.\,\ref{xmmspec}). The equivalent width (EW) of this feature is $\sim$0.25~keV.\footnote{When left free to vary, the line Gaussian width converged to $\sigma_\mathrm{E}=0.15\pm0.05$~keV, which is consistent with the pn energy resolution around 6.4~keV. So, in the fits presented in Table\,\ref{spec} we fixed the line width at zero to reduce the number of free parameters.} For the power-law fit, the reduced $\chi^2$ changes from $\chi^2_\nu=1.18$ for 52 degrees of freedom (dof) to 1.03 for 50 dof.

\begin{table*}
\begin{minipage}{17.5cm}
\centering \caption{\xmm\ and \swift\ X-ray spectroscopy. Errors are at a 1$\sigma$ confidence level for a single parameter of interest.}\label{spec}
\begin{tabular}{@{}lcccccccccc}
\hline
Obs. & Model$^{a}$ & \nh\ & $\Gamma$ & $kT$ & $R_{\mathrm{BB}}$$^{b}$ & $E$ & EW & Obs. flux$^{c}$ & Unabs. flux$^{c}$ & $\chi^2_\nu$ (dof) \\
 & & ($10^{22}$~cm$^{-2}$) & &  (keV) & (m) & (keV) & (keV) & \multicolumn{2}{c}{($10^{-12}$~\flux)} & \\
\hline
\swift/2006 & \textsc{bb} & & $2.8\pm0.5$ & $1.8^{+0.2}_{-0.1}$ & $10\pm1$ & -- & -- & $0.81\pm0.05$ & $0.98\pm05$ & 0.98~(43)\\
\swift/2006 & \textsc{pl} &  $5.1^{+0.9}_{-0.8}$ & $1.2\pm0.2$ & -- & -- & -- & -- & $0.90\pm0.05$ & $1.37^{+0.11}_{-0.09}$ & 1.14~(43)\\
\hline
\swift/2013 & \textsc{bb} & & $0.9\pm0.2$ & $2.4\pm0.2$ & $6.9^{+0.7}_{-0.6}$ & -- & -- & $0.98\pm0.05$ & $1.04\pm0.05$ & 0.98~(50)\\
\swift/2013 & \textsc{pl} &  $2.1\pm0.4$ & $0.5^{+0.2}_{-0.1}$ & --& -- &  -- & -- & $1.06\pm0.05$ & $1.22\pm0.05$ & 1.15~(50)\\
\hline
\emph{XMM} & \textsc{bb} & & $2.6^{+0.6}_{-0.4}$ & $2.2\pm0.2$ & $9\pm1$ & -- & -- & $1.19\pm0.06$ & $1.38\pm0.06$ & 1.11~(52)\\
\emph{XMM} & \textsc{bb+gau} & & $2.6^{+0.6}_{-0.4}$ & $2.1^{+0.2}_{-0.1}$ & $9\pm1$ &$6.4^{+0.2}_{-0.1}$ & $0.22^{+0.10}_{-0.07}$ & $1.18\pm0.06$ & $1.37\pm0.06$ & 1.00~(50)\\
\emph{XMM} & \textsc{pl} & $5.0^{+0.8}_{-0.7}$ & $0.9\pm0.2$ & -- & -- & -- & -- & $1.24\pm0.06$ & $1.77^{+0.12}_{-0.10}$ & 1.18~(52)\\
\emph{XMM} & \textsc{pl+gau} & $5.0^{+0.8}_{-0.7}$ & $1.0\pm0.2$ & -- & -- & $6.32^{+0.26}_{-0.05}$ & $0.25^{+0.11}_{-0.09}$ & $1.23\pm0.06$ & $1.78^{+0.12}_{-0.10}$ & 1.03~(50)\\
\hline
\end{tabular}
\begin{list}{}{}
\item[$^{a}$] \textsc{xspec} models: \textsc{bb=phabs(bbodyrad)}, \textsc{pl=phabs(powerlaw)}, \textsc{pl+gau=phabs(powerlaw+gaussian)}.
\item[$^{b}$] Blackbody radius is calculated at infinity and assuming an arbitrary distance to the source of 1~kpc. 
\item[$^{c}$] Observed (absorbed) and unabsorbed fluxes, in the 1--10 keV energy range.
\end{list}
\end{minipage}
\end{table*}

\begin{figure*}
\centering
\resizebox{\hsize}{!}{\includegraphics[angle=0]{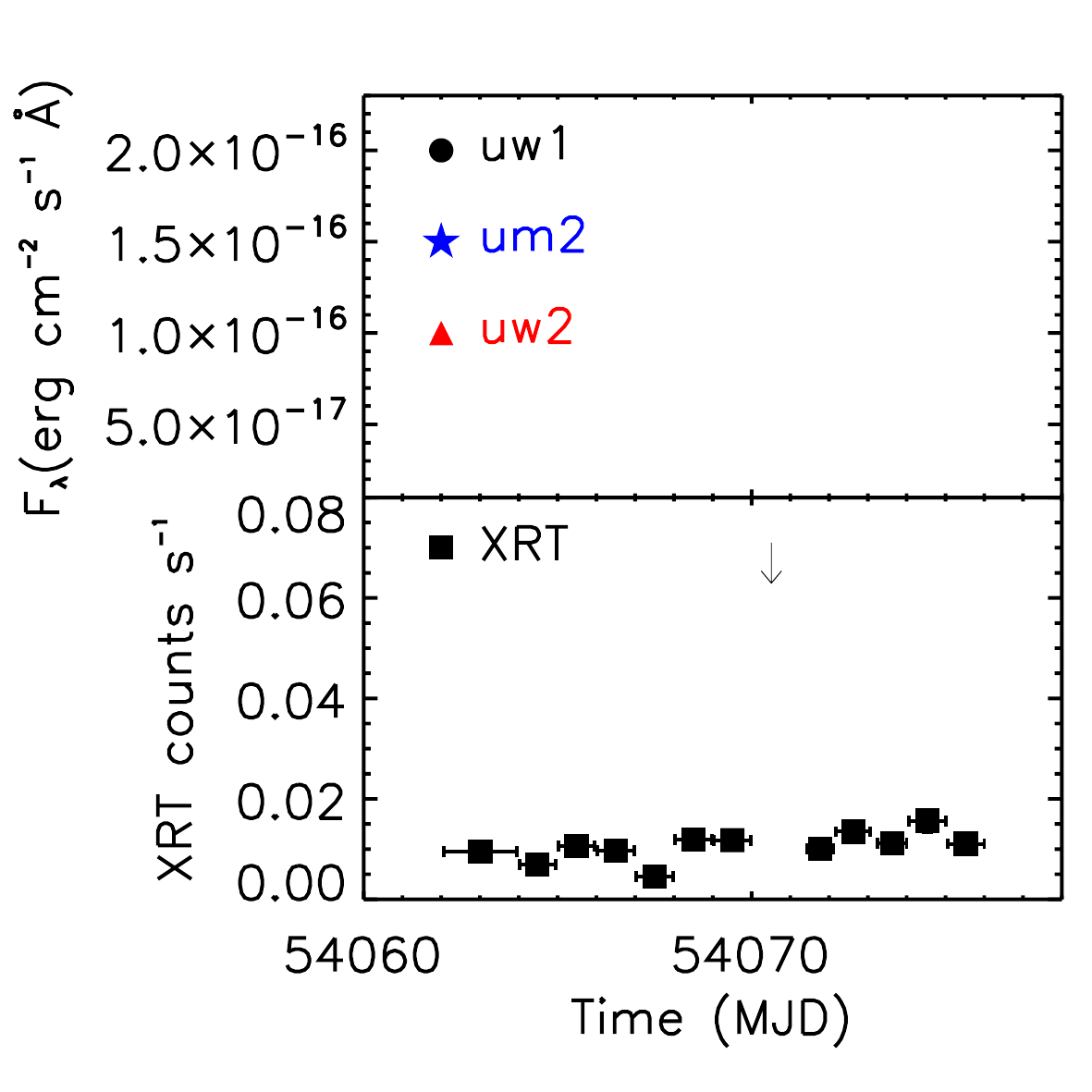}\includegraphics[angle=0]{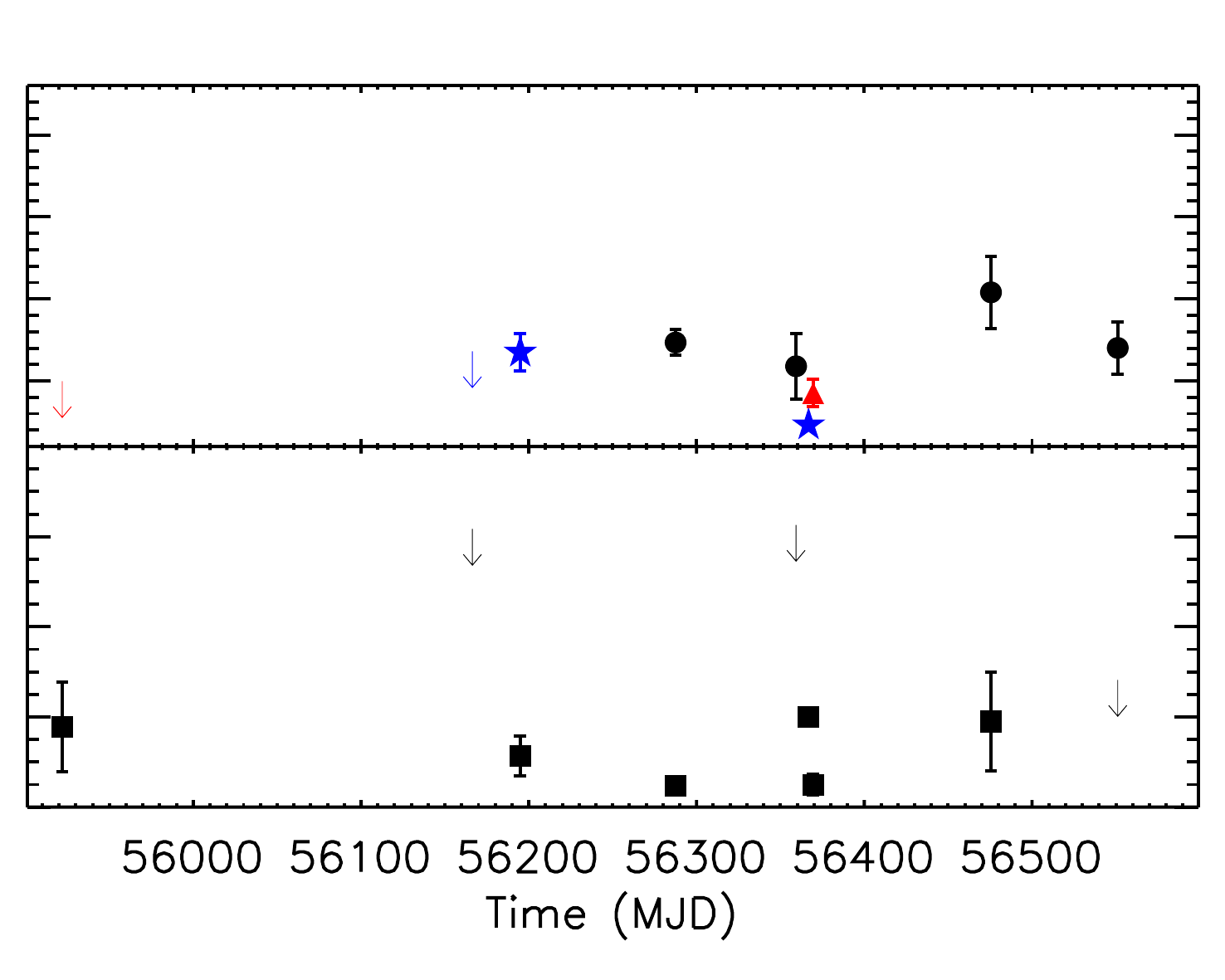}\includegraphics[angle=0]{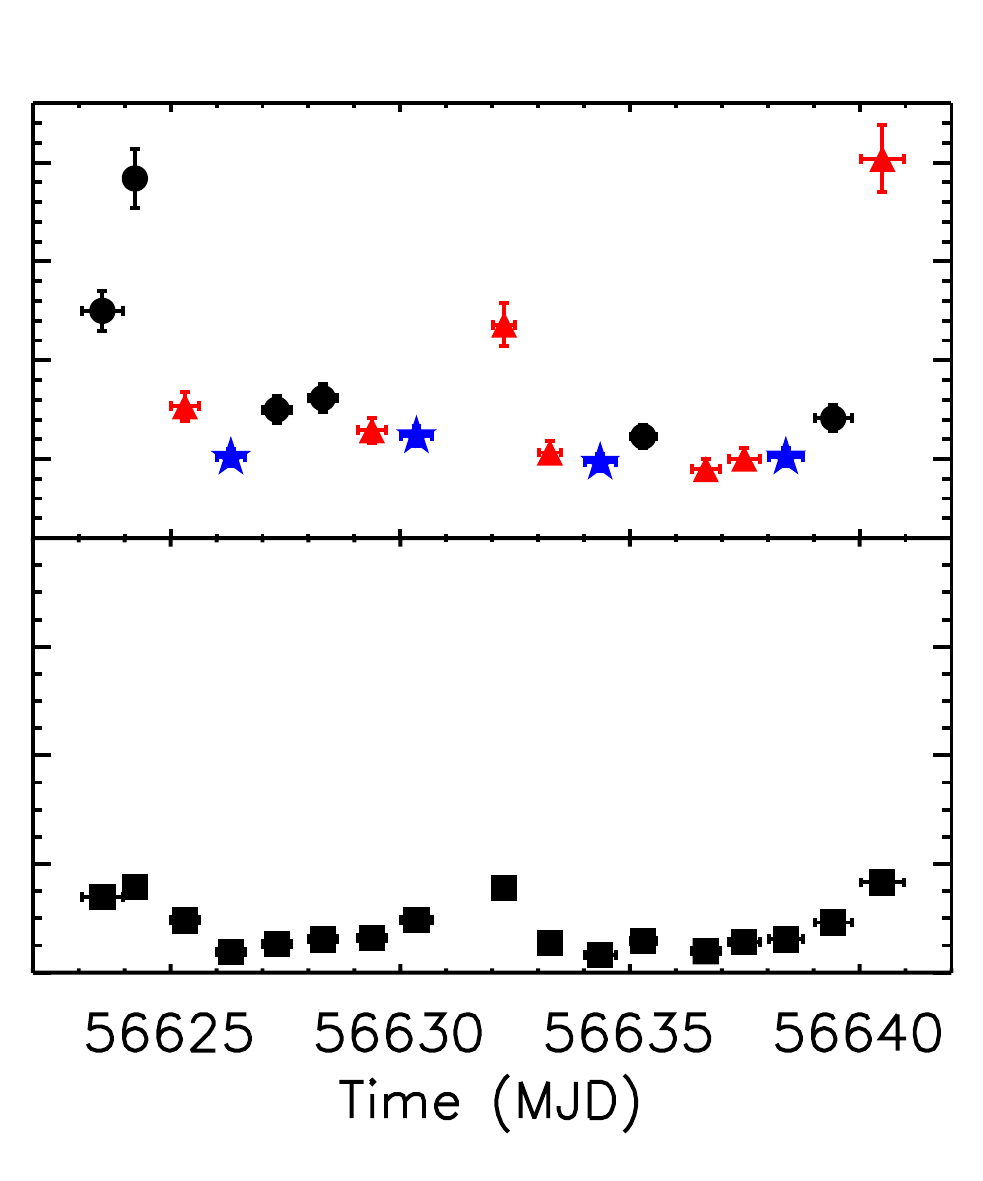}}
\caption{\label{swiftlc} Long-term \swift\ XRT and UVOT light curve of \src. Note the time gaps and the different time-scales. The down-arrows indicate upper limits at the 3$\sigma$ confidence level.}
\end{figure*}

The \swift\ count rates in the individual observations (see Table\,\ref{xrt}) show that the variability of \src\ is actually larger than between the combined XRT spectra and \xmm\ and is of a factor of $\approx$3. In Fig.\,\ref{swiftlc}, we show the long-term \swift\ XRT light curve, together with the simultaneous UVOT near-UV fluxes (when available). The X-ray and UV luminosities are roughly correlated, again suggesting the 
emissions in the different bands  are from the same body. The measured UVOT magnitudes (in the Vega photometric system, see \citealt{poole08} for more details and \citealt{breeveld10,breeveld11} for the most updated zero-points and count rate-to-flux conversion factors) are between $\sim$18.3 and 19.6 for the $uw1$ filter, 18.3 and 19.9 for the $um2$ filter, and 19.14 and 20.3 for the $uw2$ filter. In the \xmm/OM data the observed $B$ magnitude was $\sim$18.9.

\section{TNG observations}\label{tngsect}

\src\ optical spectra were obtained on 2011 July 21, August 2 and November 28 (see Table\,\ref{tab_log_spe}) with the 3.58-m TNG equipped with the Device Optimised for the LOw RESolution\footnote{See http://www.tng.iac.es/instruments/lrs/.} (DOLORES). The camera is a 2048$\times$2048 E2V 4240 Thinned back-illuminated, deep-depleted CCD with a pixel size of 13.5~$\mu$m. The field of view is 8.6$'$$\times$8.6$'$ with a 0.252~arcsec~pixel$^{-1}$ scale. We performed 900~s-long exposures on the target using the low-resolution grism LR-B, which allowed us to obtain a spectral coverage between 3000 and 8400~\AA\ and a dispersion of 2.5~\AA~pixel$^{-1}$. 
\begin{table}
\caption{Log of the TNG/DOLORES optical spectroscopy observations. \label{tab_log_spe}}
\centering
\begin{tabular}{ccc} \hline 
Spectrum &   Date obs. (mid exposure, UT) &  Exposure\\
         &  (YYYY-MM-DD)            & (ks)  \\ \hline
 1 	 &   2011-07-21.00251  	& 0.9 \\
 2 	 &   2011-08-02.09659  	& 0.9 \\
 3 	 &   2011-11-28.83850  	& 0.9 \\
 4 	 &   2011-11-28.84851  	& 0.9 \\
 5 	 &   2011-11-28.85971  	& 0.9 \\ 
 6 	 &   2011-11-28.87042  	& 0.9 \\ 
\hline
\end{tabular}   
\label{tab_log_spe}
\end{table}

Standard data reduction, including bias subtraction and flat-fielding, was performed using different packages in the Image Reduction and Analysis Facility (\textsc{iraf}) package to obtain one-dimensional flux- and wavelength-calibrated spectra. 
In particular, the spectra were wavelength-calibrated using an Ne+Hg and He reference spectrum. Corrections for atmospheric extinction were also performed. The spectro-photometric standard stars used for the relative flux calibration are: BD\,+28\,4211 (July 21 and November 28 spectra), Feige\,110 (August 2 spectrum). The log of spectroscopic observations is reported in Table~\ref{tab_log_spe}.

\begin{figure*}
\centering
\resizebox{\hsize}{!}{\includegraphics[angle=-90]{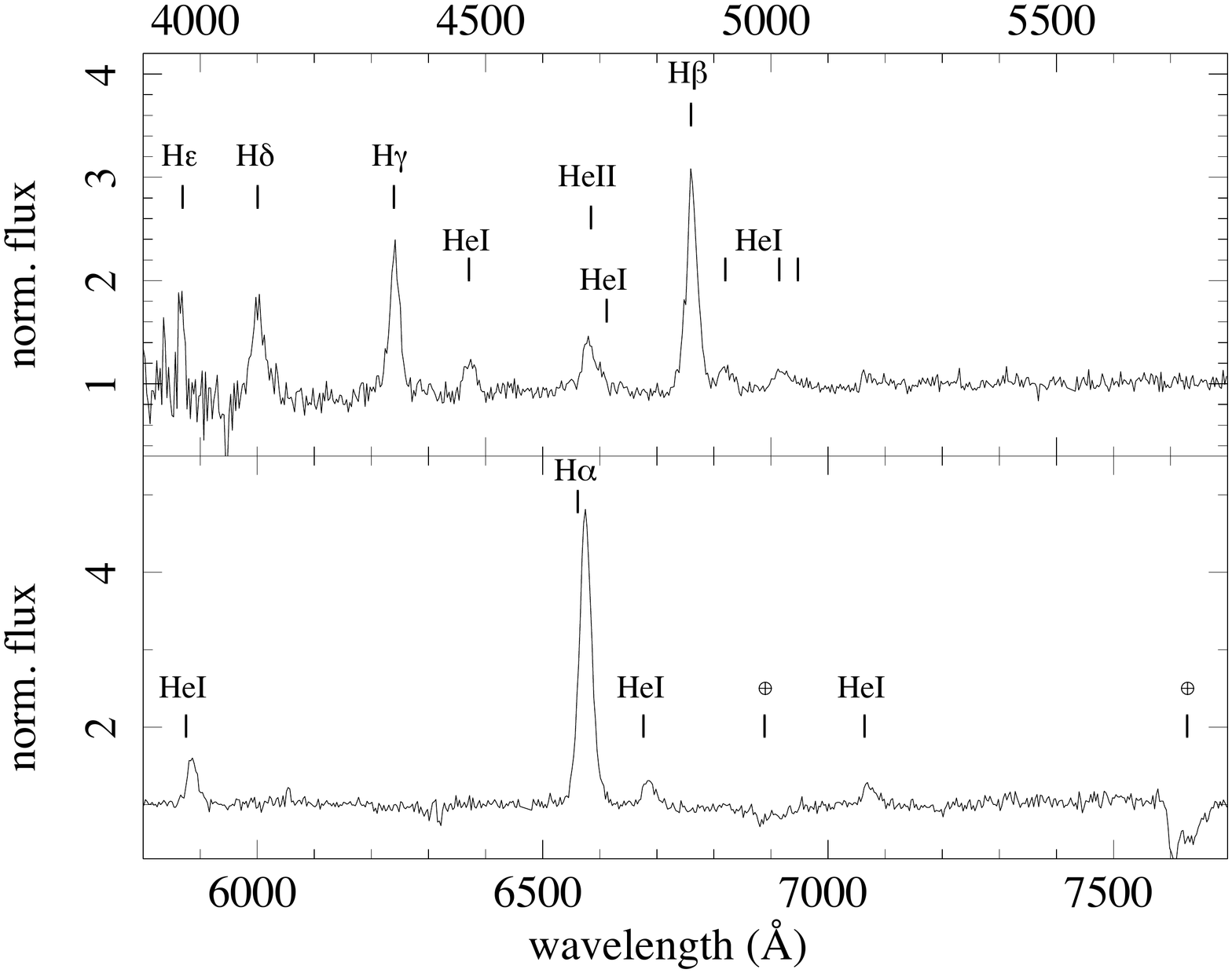}}
\caption{\label{tngspec} TNG spectra of \src. The identified lines as well as the main atmospheric absorption bands are labelled.}
\end{figure*}

The flux-calibrated optical spectra of \src\ display a blue continuum with broad emission lines superposed. We clearly detect H$\alpha$, H$\beta$, H$\gamma$, H$\delta$, He\textsc{i} ($\lambda\lambda$ 4472, 4713, 4921, 5016--5048, 5876, 6678, 7065 \AA), and He\textsc{ii} ($\lambda 4686$). No clear absorption feature is detected. The continuum-normalized average spectrum is shown in Fig.\,\ref{tngspec}. 

The presence of broad (FWHM $\sim 1000$--2000~km~s$^{-1}$) H and He emission lines favour an origin in an accretion flow on to a compact object, either an NS or a WD.
Furthermore, some emission lines show a hint for a double-horned profile, which might be related to the presence of an accretion disc. 
Several lines show significant intensity variations. The EWs and FWHM for the main detected emission lines obtained by Gaussian fits to the emission profiles are reported in Table~\ref{lines}. 

Radial velocities were measured for the four consecutive spectra obtained during the night of 2011 November 28 (spectra 3--6 of Table~\ref{tab_log_spe}). Relative Doppler shifts were measured with respect to the first spectrum of the series, taken as reference, trough cross-correlation of the main emission lines in the wavelength range 4250--6800~\AA. The results are shown in Table~\ref{tab_rad_vel}. 

Finally, we note that there might be some variability in the source fluxes observed by TNG. In particular, in the night of 2011 November 28 (Table\,\ref{tab_log_spe}), when four observations were carried out within hours, the fluxes were $6.7\times10^{-14}$, $7.8\times10^{-14}$, $8.2\times10^{-14}$, and $7.6\times10^{-14}$~\flux\ for the four spectra in chronological order (we estimate an uncertainty on the fluxes of 20\%). These values are consistent with each other within the uncertainties, but the nominal variability, if true, could be due to the orbital motion, and also effects of the spin rotation cannot be  excluded.

\begin{table}
\caption{Results of optical emission lines analysis.\label{lines}}
\centering
\begin{tabular}{ccc} \hline 
Spectrum       &  EW                &  FWHM               \\
               & (\AA)                & (km s$^{-1}$)         \\ \hline
\multicolumn{3}{|c|}{H$\delta$}                           \\ \hline
 1 	       & $  -3.4 \pm 1.4$   &  $2259 \pm  446$	  \\
 2 	       & $ -11.4 \pm 2.3$   &  $1667 \pm  204$	  \\
 3 	       & $ -11.3 \pm 4.6$   &  $2049 \pm  775$	  \\
 4 	       & $ -15.0 \pm 4.0$   &  $2610 \pm 1718$	  \\
 5 	       & $ -13.3 \pm 5.7$   &  $2186 \pm  738$	  \\
 6 	       & $  -           $   &  $2932 \pm 2303$    \\
\hline
\multicolumn{3}{|c|}{H$\gamma$}         	          \\ \hline
 1 	       & $  -5.6 \pm 1.4$   &  $2031 \pm 276$	  \\
 2 	       & $ -25.7 \pm 2.1$   &  $1368 \pm 110$	  \\
 3 	       & $ -25.0 \pm 3.8$   &  $1347 \pm 165$	  \\
 4 	       & $ -16.1 \pm 3.0$   &  $1485 \pm 241$	  \\
 5 	       & $ -10.7 \pm 3.7$   &  $1561 \pm 214$	  \\
 6 	       & $  -9.8 \pm 4.1$   &  $1326 \pm 241$	  \\
\hline
\multicolumn{3}{|c|}{H$\beta$}               	          \\ \hline
 1 	       & $-12.9 \pm  0.9$   &  $1852 \pm 204$	  \\
 2 	       & $-57.5 \pm  1.2$   &  $ 124 \pm  74$	  \\
 3 	       & $-61.5 \pm  1.9$   &  $1179 \pm 117$	  \\
 4 	       & $-45.8 \pm  1.2$   &  $1333 \pm  86$	  \\
 5 	       & $-52.1 \pm  1.2$   &  $1568 \pm 117$	  \\
 6 	       & $-43.8 \pm  1.6$   &  $1580 \pm  99$	  \\
\hline
\multicolumn{3}{|c|}{H$\alpha$}               	          \\ \hline
 1 	       & $  -7.9 \pm 2.5$   &  $ 950 \pm 192$     \\
 2 	       & $-119.2 \pm 1.5$   &  $1049 \pm  32$     \\
 3 	       & $-127.5 \pm 2.2$   &  $ 905 \pm  41$     \\
 4 	       & $-120.5 \pm 1.8$   &  $1087 \pm  32$     \\
 5 	       & $-125.3 \pm 1.4$   &  $1220 \pm  32$     \\
 6 	       & $-107.5 \pm 2.3$   &  $1295 \pm  32$     \\ \hline
 \multicolumn{3}{|c|}{He\textsc{i} 4472}  	     	          \\ \hline
 1 	       & $ - $              &  $1375 \pm 1053$    \\
 2 	       & $  -5.9 \pm 1.3$   &  $1677 \pm  476$    \\
 3 	       & $  -8.2 \pm 2.0$   &  $1113 \pm  394$    \\
 4 	       & $ -11.9 \pm 1.6$   &  $2012 \pm  550$    \\
 5 	       & $  -8.4 \pm 1.9$   &  $1489 \pm  442$    \\
 6 	       & $ -10.6 \pm 2.0$   &  $1522 \pm 1006$    \\
\hline
\end{tabular}   
\end{table}
\begin{table}
\contcaption{Results of optical emission lines analysis.}
\centering
\begin{tabular}{ccc} \hline 
Spectrum       &  EW                &  FWHM               \\
               & (\AA)                & (km s$^{-1}$)         \\ \hline

\multicolumn{3}{|c|}{He\textsc{i} 5876}  	     	          \\ \hline
 1 	       & $ - $              &  $ -          $     \\
 2 	       & $ -21.7 \pm 1.7$   &  $ 873 \pm  97$     \\
 3 	       & $ -20.1 \pm 2.5$   &  $ 959 \pm 144$     \\
 4 	       & $ -17.6 \pm 2.0$   &  $1082 \pm 144$     \\
 5 	       & $  -8.0 \pm 2.2$   &  $1296 \pm 168$     \\
 6 	       & $ -12.1 \pm 2.4$   &  $1322 \pm 178$     \\
\hline
\multicolumn{3}{|c|}{He\textsc{i} 6678}  	     	          \\ \hline
 1 	       & $  -2.0 \pm 1.6$   &  $ 485 \pm 359$	  \\
 2 	       & $ -10.6 \pm 1.1$   &  $1295 \pm 220$     \\
 3 	       & $  -5.2 \pm 1.3$   &  $1412 \pm 296$     \\
 4 	       & $  -8.1 \pm 1.1$   &  $1057 \pm 251$     \\
 5 	       & $  -6.1 \pm 1.0$   &  $1902 \pm 517$     \\
 6 	       & $  -7.5 \pm 1.3$   &  $1511 \pm 814$	  \\
\hline
\multicolumn{3}{|c|}{He\textsc{i} 7065}  	     	          \\ \hline
 1 	       & $  -3.0 \pm 1.4$   &  $322 \pm 106$      \\
 2 	       & $ -            $   &  $ - $              \\
 3 	       & $  -7.7 \pm 1.3$   &  $ 569 \pm 219$	  \\
 4 	       & $  -5.9 \pm 1.2$   &  $1397 \pm 369$	  \\
 5 	       & $  -3.9 \pm 1.0$   &  $ 985 \pm 318$	  \\
 6 	       & $  -5.7 \pm 1.5$   &  $2263 \pm 598$	  \\
\hline
\multicolumn{3}{|c|}{He\textsc{ii} 4686}  	     	          \\ \hline
 3-6$^a$ 	& $  -8.2 \pm 0.6$  &  $1325 \pm 211$     \\
\hline
\end{tabular}   
\begin{list}{}{}
\item[$^{a}$] Values obtained from the average of the spectra 3, 4, 5, and 6.
\end{list}
\end{table}

\begin{table}
\caption{Radial velocities of the main spectral emission lines cross-correlating the November 28 spectra in the wavelength range
4250--6800~\AA. Spectrum no.~3 was used as reference.
}
\centering
\begin{tabular}{ccc} \hline 
Spectrum &   Date obs. (mid exposure) &  Radial velocity   \\
         &   YYYY-MM-DD                 &  km s$^{-1}$	   \\ \hline
 3 	 &   2011-11-28.83850  	      &  --  \\
 4 	 &   2011-11-28.84851  	      &  $-12.3 \pm 10.8$  \\
 5 	 &   2011-11-28.85971  	      &  $ -6.4 \pm  7.5$  \\ 
 6 	 &   2011-11-28.87042  	      &  $ 96.8 \pm 19.4$  \\ 
\hline
\end{tabular}   
\label{tab_rad_vel}
\end{table}

\section{Loiano observations}\label{loiano}
Optical photometric data were secured during the night of 2011 September 20 at the Cassini 1.52-m Telescope in Loiano (Italy) using the Bologna Faint Object Spectrograph \& Camera\footnote{See http://davide2.bo.astro.it/wp-content/uploads/2013/09/rep04-2001-01-textfig.pdf.} (BFOSC). The data set consists of 200 images acquired in the $V$ filter in windowed mode, i.e. selecting a central box of 670$\times$355 pixel, to improve the readout speed. The images have an exposure time of 80~s each and were obtained during a 4.5~h time interval from UT 18:54 to UT 23:31. The sky conditions were clear, although non-photometric, and the average seeing was around $2\farcs5$.

The images were processed using standard procedures and bias and flat-field frames acquired in the same night. All the images were carefully registered and co-added to generate a master frame that was used to obtain a master list of objects. This list was then used as input for the \textsc{daophot} pipeline in the \textsc{iraf} package, and point spread function photometry, as well as aperture correction, were obtained. 
The magnitudes were then reported to one of them chosen as master (frame \#117) because of its good overall quality (good seeing and ellipticity, resulting in a smaller scatter in a plot of instrumental magnitude versus internal photometric error). Information on the target and five comparison stars was then extracted and a light curve was created. 
Since the night was non-photometric, no photometric standards were available. Moreover, this portion of sky is not included in almost any optical survey. Therefore, we decided to calibrate our $V$ instrumental magnitudes by using the photographic $V$ magnitude in the Guide Star Catalog ver.~2.3 \citep[GSK\,2.3;][]{lasker08}, which has many stars in our field of view (the same catalogue was also used to improve the astrometry). The calibration is not extremely precise also because we have only one band and thus no colour term correction is possible. 
We derived a zero-point ${\mathrm{z.p.}_{V} = 23.1 \pm 0.1}$.

\begin{figure}
\centering
\resizebox{\hsize}{!}{\includegraphics[angle=0]{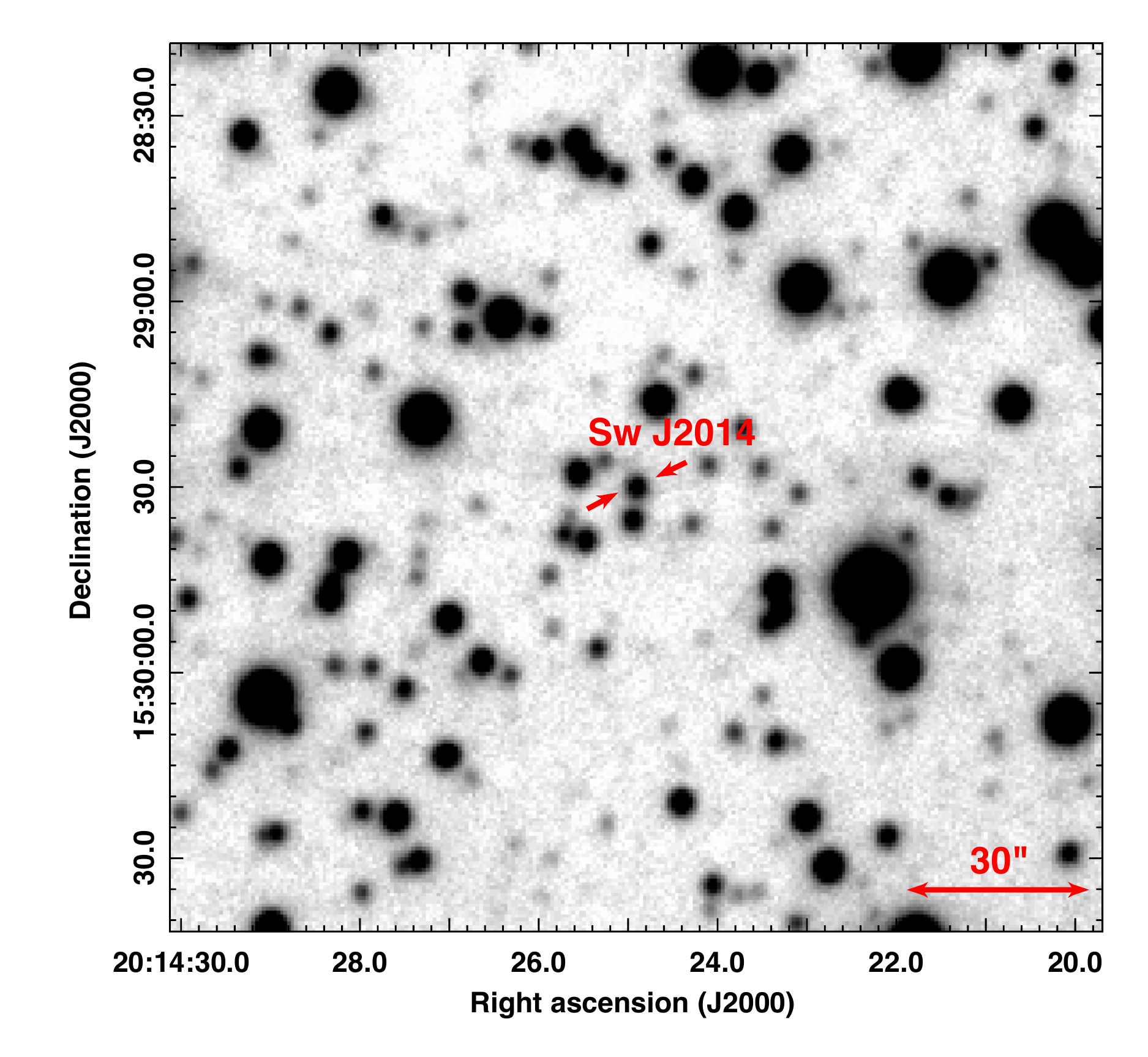}}
\caption{\label{fchart} Loiano $B$-band image of the field of \src\ (marked and labelled). The net exposure was 80~s. The image side is 2.5 arcmin wide.}
\end{figure}

Only one optical source in the BFOSC images is consistent with the \xmm\ position of \src, even allowing for a generous error radius of 2~arcsec (see Fig.\,\ref{fchart}). Best-fitting position (J2000): $\rm RA=20^h14^m24\fs91,~Dec.=+15\degr29'30\farcs0$ (J2000, not heliocentric corrected); the uncertainty is dominated by the seeing ($\sim$1$''$). The position of the optical counterpart of \src\ is  consistent with those reported in the USNO A2 and B1 catalogues. In particular we identify it with  the USNO A2 star U1050-16948097  at  $\rm RA=20^h14^m24\fs98,~Dec.=+15\degr29'30\farcs6$, reported in the catalogue at $R=18.8$ and $B=18.4$. It is also identified with the older USNO B1 star 1054-0559015 at $\rm RA=20^h14^m24\fs9,~Dec.=+15\degr29'30\farcs5$ reported with $B=19.3$, $R=18.7$ and $I=18.5$. 

\begin{figure}
\centering
\resizebox{\hsize}{!}{\includegraphics[angle=-90]{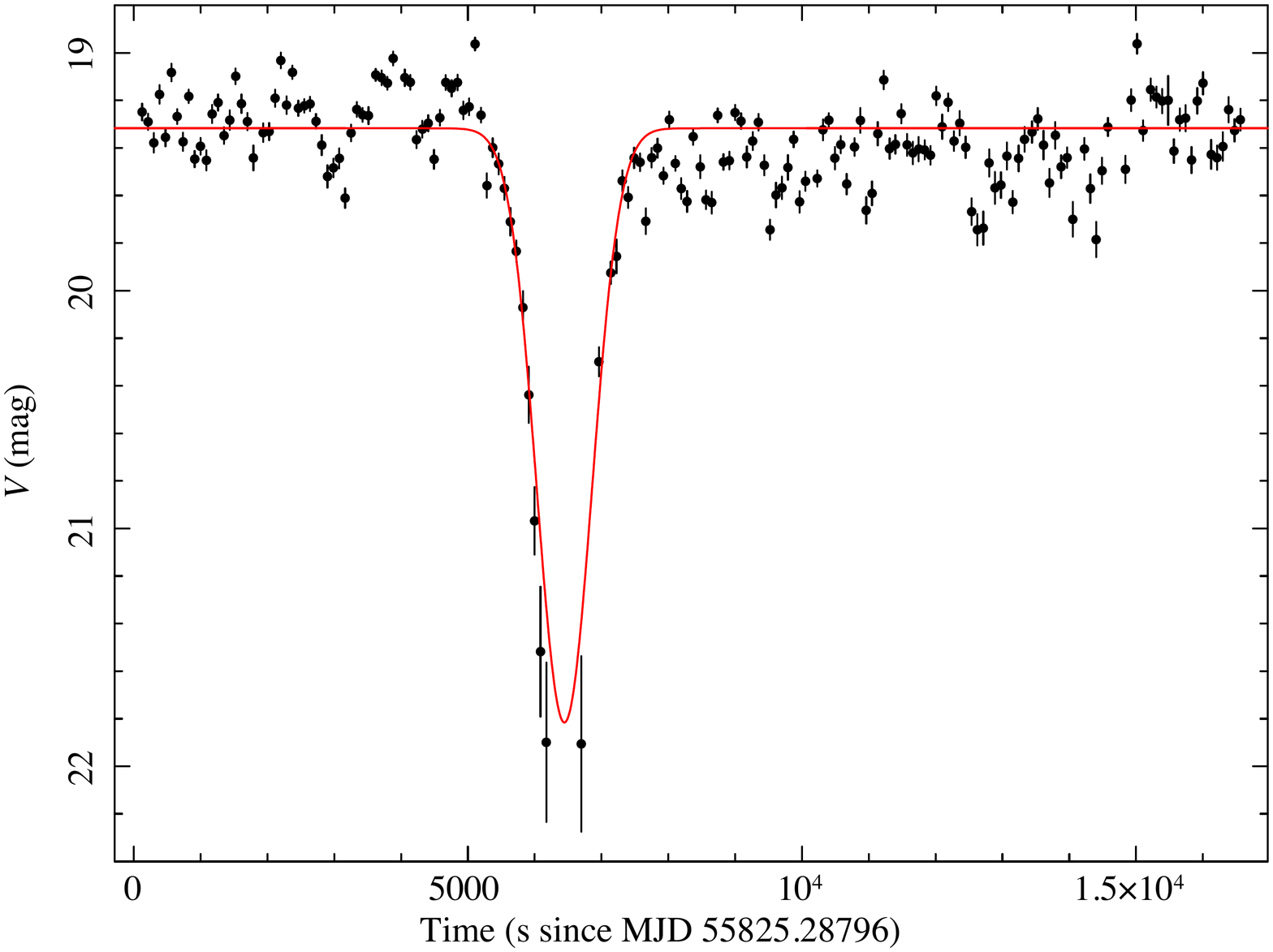}}
\caption{\label{loianolc} Loiano $V$-band light curve of \src. The red, solid line indicates the constant-plus-Gaussian model used to infer the time and the duration of the eclipse.}
\end{figure}

Figure\,\ref{loianolc} shows the $V$-band Loiano light curve of the \src's optical counterpart.
There is a clear evidence for an eclipsing event lasting about 35 min.
The eclipse can be modelled by a Gaussian centred at 
MJD 55\,825.36254(9) and with $\sigma=414\pm10$~s ($6.90\pm0.17$~min). The FWHM duration of the eclipse is $973\pm22$~s ($16.2\pm0.4$~min). These values are consistent with the ones obtained from the X-ray eclipse.
From the light curve, it is not possible to tell whether a flat bottom is present in the eclipse. A fit with a piecewise linear model to the data constrained the duration of the flat-bottom part -- if present -- at $<$0.4~ks (at 3$\sigma$), while the inferred ingress and egress lengths are $0.5\pm0.1$~ks. We searched the light curve for the 491 and 12\,379-s modulations, but none could be found, with a 3$\sigma$ upper limit of 0.03~mag on both signals.

\section{Discussion}\label{discussion}

We have presented X-ray and optical data of the  previously unknown X-ray source SwJ2014. Its X-ray position has been determined with \xmm\ at the J2000 coordinates $\rm RA=20^h14^m24\fs96,~Decl.=+15\degr29'29\farcs04$, with an uncertainty of $\sim$$1\farcs5$. This allowed us to identify the optical counterpart of \src\ with the NOMAD source 1054--0592809 ($B\sim19.3,~R\sim18.9$), at the catalogued position $\rm RA=20^h14^m24\fs899,~Decl.=+15\degr29'30\farcs51$.
We detected X-ray pulsations at 491~s, which indicate that \src\ harbours a compact star, either a WD or a NS. The pulses are sinusoidal, with a substantial pulsed fraction (40--50\%) and there is some hint of a decreasing trend of the pulsed fraction with increasing photon energy.
The presence of an additional feature in the X-ray light curves, a permanent eclipse, is a clear signature of an accreting binary with high inclination, $i\geqslant 80\degr$. The X-ray eclipses are total, with a duration of about 20~min, and eclipses are also observed in the optical band, with similar duration. This  allowed us to determine an orbital period of 3.44~h for the system. The optical spectrum is dominated by typical accretion features, which indicate the presence of an accretion disc. It is then conceivable that the  eclipses in the two bands are related to the eclipse of the compact star and its  accretion regions by the donor star, which is undetectable in our optical spectra. The short orbital period would indicate a late type companion of spectral type M4, according tho the orbital period--spectral type relation by \citet{smith98}.

The stability of the period over a time-span of several years suggests a WD.
In fact, due to their larger momentum of inertia, WDs are much harder to spin-up or spin-down than NSs (e.g. \citealt{patterson94,bildsten97}).
 A magnetic CV of the intermediate polar (IP) type\footnote{See e.g. the catalogue\\ http://asd.gsfc.nasa.gov/Koji.Mukai/iphome/catalog/members.html.} is therefore the favoured interpretation for \src. Most IPs have spin to orbit period ratio in the range 0.25--0.01 clustering around  0.1 ( \citealt*{nws04}; see also \citealt{bernardini12}).
The spin to orbit period ratio is 0.04, indicating a rather desynchronized magnetic WD. The orbital period locates this system above the 2--3~h orbital period gap (see e.g. \citealt*{knigge11}), where most of the IPs are found. A CV is also favoured by the Balmer and He optical spectral lines. The EW of the H$\beta$ is 45.6~\AA\ (averaged) and that of He\textsc{ii}~4686 is 8.2~\AA. The ratio of their EWs are typical of CVs rather than NS systems \citep{vanparadijs84}. Moreover, the ratio of the average X-ray to  optical fluxes is about 14, which is again more typical for CVs rather than NS binaries, which generally have X-ray fluxes larger by one to two order of magnitudes \citep{warner03}.
 Also, the orbital X-ray light curve is reminiscent of those observed in IPs (e.g. \citealt*{parker05}), and the possible trend of decreasing pulsed fraction with increasing energy in the spin pulse profile, if real, would be another fairly common feature among IPs (e.g. \citealt*{rosen88,norton89}).\footnote{A few IPs are known to display spin modulations that are not energy dependent (e.g. \citealt{allan98,hellier98,demartino05}). These systems have orbital periods below the gap and likely are accreting at low rates.} These characteristics observed in IPs are signature of photoelectric absorption, being the former produced by material at the impact region at the disc rim and the latter by dense material in the magnetically channeled accretion flow.

The X-ray spectrum is hard ($\Gamma \sim 1$) and highly absorbed ($N_{\rm H}\sim5\times10^{22}$~cm$^{-2}$) indicating the presence of dense material close to the compact star. A blackbody fit gives an unacceptable temperature if the X-ray emission arises from a WD. Also the radius derived is unfeasible, since it is too small (few metres) for both WD and NS, even assuming a large distance. An optically thin component is unconstrained therefore preventing direct comparisons with X-ray spectra of IPs, which are characterized by heavily absorbed multi-temperature plasma  (e.g. \citealt*{ishida94,beardmore00}) and in many cases by a soft  blackbody component with temperatures of tens of eV (e.g. \citealt{demartino04,anzolin08,anzolin09}). 
On the other hand, X-ray spectra of quiescent NS low mass, short-period systems are characterized by a soft blackbody component arising  from the NS atmosphere but also with lower temperature than that derived. Higher temperatures up to 3~keV are generally detected during X-ray bursts (e.g. \citealt{galloway08}).

Another piece of evidence in favour of a CV classification is the value of the spin period itself.
X-ray binaries (XRBs) containing a NS (NS--XRBs) are usually classified based on the spectral type of the donor, for evolutionary reasons: NS--XRBs with late type companions are named low mass X-ray binaries (LMXBs), while NS--XRBs hosting massive early type stars are called high mass X-ray binaries.
A third class of (rare) accreting NS--XRBs has been more recently recognized, thanks to the growing number of members  (e.g. \citealt{corbet08}). They are the so-called symbiotic (NS) binaries (SyXBs). Although some of them are known since the early times of X-ray astronomy (e.g. GX\,1+4; \citealt*{lewin71}), and were at first classified as LMXBs, SyXBs display completely different properties (see e.g. \citealt{mop06,corbet08}), like slow NS pulsations, M-type giant companions and very wide orbits (for example, GX\,1+4, has P$_{\rm spin}$=114~s and P$_{\rm orb}$=1\,160.8~d). On the contrary, LMXBs where pulsations have been detected display NS spin periods typically in the millisecond range, reaching, at maximum, a spin period of $\sim$7.7~s in the LMXB 4U\,1626--67 (\citealt*{liu07}; see also figs.~1 and 13 in \citealt{enoto14}).
Given the typical properties characterizing LXMBs and SyXBs, the two periodicities we discovered in \src\ (spin period of 491~s and orbital period of 3.44~h) make  an association with an XRB containing a NS as compact object very unlikely. Indeed, this would place the source in a completely empty portion of the so-called Corbet diagram of the NS spin period versus the orbital period of the system (see an updated version of this diagram in fig.\,13 of \citealt{enoto14}).
In conclusions, while the X-ray spectra cannot constrain the type of source, all the other pieces of evidence indicate that \src\ hosts a WD, and is thus a magnetic CV of the IP type.

In the hypothesis that \src\ consists of a WD accreting from a Roche lobe-filling, low-mass, main-sequence star, with standard considerations (e.g. \citealt{paczynski71}), making the approximations of a circular orbit and $\sin i\simeq1$, and assuming standard stellar mass--radius relations (e.g. \citealt{smith98,knigge06}), we can estimate the donor mass and radius as $M_\bigstar\approx0.3~\mathrm{M_{\sun}}$ and $R_\bigstar\approx0.4~\mathrm{R_{\sun}}$. If for the WD mass we take 0.8~$\mathrm{M_{\sun}}$, which is a reasonable value in CVs \citep*{zorotovic11}, we get from Kepler's third law an estimate for the binary separation of $a\approx1.2~\mathrm{R_{\sun}}$. Considering the $\sim$0.2~ks ingress/egress duration, the linear size of the X-ray-emitting region should be of the order of $\approx$$0.05~\mathrm{R_{\sun}}$ (for $i=80\degr$; see equation\,1 of \citealt{nucita09}; see also \citealt{wheatley03}). This is $\sim$5 times the radius that can be expected for a 0.8~$\mathrm{M_{\sun}}$ WD, but considered the large uncertainties the value is consistent with the X-rays being emitted from the WD or close to it, and also with estimates of the inner radius of the truncated disc of a CV (e.g. \citealt{mhlahlo07}; \citealt*{dobrotka14}). Future multiband observations should be able to provide tight constraints on the system parameters.\\
\indent We finally notice that, assuming $1.8\times10^{-12}$~\flux\ for the unabsorbed X-ray flux of \src\ (Table\,\ref{spec}), the X-ray luminosity of the source is $L_{\mathrm{X}}\simeq2\times10^{32}d^2_1$~\lum, where $d_1$ is the distance in units of 1~kpc. For a few kpc, this is consistent with typical values for IPs ($10^{32}$--$10^{34}$~\lum; e.g. \citealt{sazonov06}).

\section{Summary}\label{summary}

In this paper, we reported on the SATS\,@\,BAR discovery of the new X-ray pulsator \src. The periodicity of 491~s indicates that \src\ harbours a compact star. Follow-up observations to investigate the nature of the source were done both with space-borne (\swift\ and \xmm) and ground-based (TNG and Loiano Telescope) facilities.\\
\indent The \xmm\ EPIC and OM light curves both show a sudden drop of the flux followed by a quick recovery, similar to the deep/total eclipses in NS--LMXBs or CVs with high inclination. The time lag between the X-ray and the optical eclipses is $\Delta t_{\rm eclipses}\simeq12.4$~ks. Unfortunately, the EPIC and OM data are not simultaneous, but the \swift\ monitoring campaign provided us evidence indicating that $\Delta t_{\rm eclipses}$ is the orbital period of the system, and thus that the X-ray and optical eclipses are simultaneous (and that the $491$-s periodicity is the rotation period of the compact object).
The X-ray spectra are quite hard (with photon index $\Gamma{}\sim{}0.8$). Optical spectra allowed us to measure several emission lines, which are consistent with an accretion disc. Photometry led to the identification of the optical counterpart (the NOMAD star 1054--0592809).\\ 
\indent The nature of the compact star (i.e. whether it is a NS or a WD) cannot be definitely assessed by our data but, in the case of a NS, \src\ would lie in a completely empty portion of the Corbet diagram for spin versus orbital period. On the other hand, several pieces of information (including the long spin period and its stability over a time-span of several years, the EWs and the ratio of the H$\beta$ and He\textsc{ii} 4686) point towards a WD.
We are thus confident that \src\ is a magnetic CV of the IP type. In particular, \src\ is a potential addition to the very small number of known deeply-eclipsing IPs  (\citealt*{walker56,hellier97,hmb97}; \citealt{woudt03,warner09,aungwerojwit12}). These systems are very important since they allow  accurate measurements of the system parameters and the possibility to study the geometry of the X-ray emitting regions (see e.g., \citealt{hellier14}).

\section*{Acknowledgements} 
This research is based on observations obtained with \swift\ and \xmm. \swift\ is a NASA mission with participation of the Italian Space Agency and the UK Space Agency. \xmm\ is an ESA science mission with instruments and contributions directly funded by ESA Member States and NASA. We also made use of observations collected at the Astronomical Observatory of Bologna in Loiano and of TNG operated on the island of La Palma by the Fundaci\'on Galileo Galilei of the INAF. The \textsc{iraf} software package is distributed by the NOAO, which is operated by AURA, Inc., under cooperative agreement with the NSF. We thank the anonymous referee for valuable comments on the manuscript and constructive suggestions. PE acknowledges a Fulbright Research Scholar grant administered by the US--Italy Fulbright Commission and is grateful to the Harvard--Smithsonian Center for Astrophysics for hosting him during his Fulbright exchange. (The Fulbright Scholar Program is sponsored by the US Department of State and administered by CIES, a division of IIE.) MM acknowledges financial support from the Italian Ministry of Education, University and Research (MIUR) through grant FIRB 2012 RBFR12PM1F, and from INAF through grant PRIN-2011-1. This work was partially supported by the following contracts and grants: ASI/INAF I/004/11/0 and I/037/12/0, SAO/CXC~GO3-14093X, NSF~AST-1211843, and NASA~NNX12AE39G.

\bibliographystyle{mn2e}
\bibliography{biblio}

\bsp

\label{lastpage}

\end{document}